\documentclass[conference,review,anonymous]{IEEEtran}
\usepackage{cite}
\usepackage{amsmath,amssymb,amsfonts}
\usepackage{lipsum}
\usepackage{graphicx}
\usepackage{textcomp}
\usepackage{xcolor}
\usepackage{comment}
\usepackage{acronym}
\def\BibTeX{{\rm B\kern-.05em{\sc i\kern-.025em b}\kern-.08em
    T\kern-.1667em\lower.7ex\hbox{E}\kern-.125emX}}
\usepackage[frozencache=true,cachedir=minted-cache]{minted}

\newif\ifDEBUG
\DEBUGfalse

\newif\ifANONYMOUS
\ANONYMOUSfalse

\newif\ifJOURNAL
\JOURNALfalse

\usepackage{booktabs} 
\usepackage{amsmath}
\usepackage{algorithm}
\usepackage{algpseudocode}
\usepackage[normalem]{ulem} 
\usepackage{xspace}
\usepackage{multirow} 
\usepackage{balance} 
\usepackage{fancyvrb} 
\usepackage{caption}

\usepackage{enumitem}
\setlist[itemize]{leftmargin=*,noitemsep,topsep=0pt}
\setlist[enumerate]{leftmargin=*}

\usepackage{etoolbox}
\makeatletter
\patchcmd{\@makecaption}
	{\scshape}
	{}
	{}
	{}
\makeatletter
\patchcmd{\@makecaption}
	{\\}
	{.\ }
	{}
	{}
\makeatother
\def\tablename{Table}

\newcommand{\code}[1]{\texttt{{\small #1}}}
\newcommand{\inlinecode}[1]{%
  \mintinline[fontsize=\footnotesize{},mathescape, escapeinside=||]{cpp}{#1}%
}

\newcommand{\ie}{\textit{i.e.,}\xspace}
\newcommand{\eg}{\textit{e.g.,}\xspace}
\newcommand{\etal}{\textit{et al.}\xspace}

\usepackage{mdframed}
\mdfsetup{skipabove=0.5\topskip,skipbelow=0.5\topskip,align=center}

\usepackage{amsthm}
\newtheorem{thm}{Theorem}

\newenvironment{smalldescription}{
	 \setlength{\topsep}{0pt}
	 \setlength{\partopsep}{0pt}
	 \setlength{\parskip}{0pt}
	 \begin{description}[style=unboxed]
	 \setlength{\leftmargin}{1in}
	 \setlength{\parsep}{0pt}
	 \setlength{\parskip}{0pt}
	 \setlength{\itemsep}{0pt}
	 }
	 {\end{description}}

\DeclareMathSymbol{\mlq}{\mathord}{operators}{``}
\DeclareMathSymbol{\mrq}{\mathord}{operators}{`'}

\newif\ifSAVESPACE
\SAVESPACEfalse

\ifSAVESPACE

\else

\fi

\usepackage{todonotes}
\usepackage{soul}
\usepackage{fontawesome}

\ifDEBUG
    \newcommand{\AH}[1]{\todo[color=cyan,inline]{AH:#1}}
    \newcommand{\JD}[1]{\todo[color=yellow,inline]{JD:#1}}
    \newcommand{\FS}[1]{\todo[color=green,inline]{FS:#1}}
    \newcommand{\PA}[1]{\todo[color=orange,inline]{PA:#1}}
    \newcommand{\KR}[1]{\todo[color=yellow,inline]{Kyle:#1}}
    
    \newcommand{\TODO}[1]{\hl{#1}}
\else
    \newcommand{\AH}[1]{}
    \newcommand{\JD}[1]{}
    \newcommand{\FS}[1]{}
    \newcommand{\PA}[1]{}
    \newcommand{\KR}[1]{}
    
    \newcommand{\TODO}[1]{#1}
\fi

\usepackage{hyperref}

\usepackage{cleveref}
\crefformat{section}{\S#2#1#3}
\crefname{figure}{Figure}{Figures}
\crefname{table}{Table}{Tables}
\crefname{theorem}{Theorem}{Theorems}
\crefname{thm}{Theorem}{Theorems}
\crefname{lemma}{Lemma}{Lemmata}
\crefname{equation}{Eqt.}{Eqts.}
\crefformat{Grammar}{Grammar #1}
\crefname{appendix}{Appendix}{Appendices}
\crefname{listing}{Listing}{Listings}

\acrodef{ENS}{Embedded Network Stack}
\newcommand{\packetdrill}{{\sc PacketDrill}\xspace}
\newcommand{\pd}{{\sc PacketDrill++}\xspace}
\newcommand{\ourdataset}{{\sc ENSBench}\xspace}

\newcommand{\myparagraph}[1]{\paragraph{#1}}
\renewcommand{\myparagraph}[1]{\vspace{0.25em} \noindent \textbf{#1:}}

\usepackage{url}

\usepackage{tcolorbox}

\makeatletter
\newcommand{\linebreakand}{%
  \end{@IEEEauthorhalign}
  \hfill\mbox{}\par
  \mbox{}\hfill\begin{@IEEEauthorhalign}
}
\makeatother

\usepackage{pifont}

\newcommand{\networkstackcount}{4\xspace}
\newcommand{\vulnerabilitycount}{61\xspace}
\newcommand{\networkstackstudiedcount}{6\xspace}
\newcommand{\vulnerabilitycountperc}{100\%\xspace}
\newcommand{\rtoscount}{6\xspace}
\newcommand{\rtosfuzzcount}{1\xspace}
\newcommand{\totalvulncount}{81\xspace}

\newcommand{\nontechnicalvulncount}{15\xspace}
\newcommand{\nonpvvulncount}{5\xspace}

\newcommand{\newvulncount}{7\xspace}
\newcommand{\freertosvulncount}{11\xspace}
\newcommand{\picotcpvulncount}{12\xspace}
\newcommand{\contikivulncount}{24\xspace}
\newcommand{\lwipvulncount}{1\xspace}
\newcommand{\zephyrvulncount}{11\xspace}
\newcommand{\fnetvulncount}{2\xspace}
\newcommand{\fuzzduration}{24\xspace}
\newcommand{\singletesttime}{XXX\xspace}
\newcommand{\multiratedlabels}{13\xspace}
\newcommand{\cohenkappa}{0.82\xspace}

\newcommand{\freertosrecreatedvuln}{5}
\newcommand{\picotcprecreatedvuln}{5}
\newcommand{\contikirecreatedvuln}{2}
\newcommand{\recreatedvulncount}{12\xspace}
\newcommand{\possiblerecreationcount}{14\xspace}

\newcommand{\freertosnettransvuln}{5}
\newcommand{\picotcpnettransvuln}{7}
\newcommand{\contikinettransvuln}{2}

\newcommand{\contikinewvuln}{2}
\newcommand{\picotcpnewvuln}{4}
\newcommand{\lwipnewvuln}{1}

\newcommand{\memoryvulnpercent}{70}
\newcommand{\lengthandpacketsizepercent}{69}
\newcommand{\netvulnpercent}{41}
\newcommand{\appvulnpercent}{29}
\newcommand{\lessthreemutationspercent}{95}
\newcommand{\statefulvulnpercent}{30}
\newcommand{\statelessvulnpercent}{70}
\newcommand{\truncatepercent}{25}

\newcommand{\onechange}{34\xspace}
\newcommand{\twochanges}{24\xspace}
\newcommand{\moretwochanges}{3\xspace}

\newcommand{\onechangeperc}{56\%}
\newcommand{\twochangesperc}{39\%}
\newcommand{\moretwochangesperc}{5\%}

\newcommand{\numdependentfields}{40\xspace}
\newcommand{\numprotocols}{15\xspace}

\newcommand{\freertossize}{42.2K\xspace}
\newcommand{\freertosstars}{3.6K (76)\xspace}
\newcommand{\freertosforks}{1.2k (110)\xspace}
\newcommand{\freertoscvediscovered}{0\xspace}

\newcommand{\contikisize}{41.6K\xspace}
\newcommand{\contikistars}{1.1K\xspace}
\newcommand{\contikiforks}{635\xspace}

\newcommand{\zephyrsize}{95.7K\xspace}
\newcommand{\zephyrstars}{7.7K\xspace}
\newcommand{\zephyrforks}{4.8K\xspace}
\newcommand{\zephyrcverecreated}{\emph{Not attempted}\xspace}
\newcommand{\zephyrcvediscovered}{\emph{Not attempted}\xspace}

\newcommand{\picotcpsize}{32.7K\xspace}
\newcommand{\picotcpstars}{1K\xspace}
\newcommand{\picotcpforks}{201\xspace}

\newcommand{\lwipsize}{84.3K\xspace}
\newcommand{\lwipstars}{525\xspace}
\newcommand{\lwipforks}{249\xspace}
\newcommand{\lwipcverecreated}{\emph{Not attempted}\xspace}

\newcommand{\fnetsize}{18.0K\xspace}
\newcommand{\fnetstars}{106\xspace}
\newcommand{\fnetforks}{46\xspace}
\newcommand{\fnetcverecreated}{\emph{Not attempted}\xspace}
\newcommand{\fnetcvediscovered}{\emph{Not attempted}\xspace}

\newcommand{\outofboundsreadtype}{22 (36\%)\xspace}
\newcommand{\outofboundswritetype}{21 (34\%)\xspace}
\newcommand{\integeroverflowtype}{5 (8\%)\xspace}
\newcommand{\integerunderflowtype}{4 (7\%)\xspace}
\newcommand{\nullpointerdereferencetype}{4 (7\%)\xspace}
\newcommand{\othertype}{5 (8\%)\xspace}
\newcommand{\totaltype}{61 (100\%)\xspace}

\newcommand{\missinglenghtfieldvalidation}{23 (38\%\xspace}
\newcommand{\missingpacketsizevalidation}{19 (31\%)\xspace}
\newcommand{\missingheadervaluevalidation}{7 (12\%)\xspace}
\newcommand{\missingintegerwraparoundvalidation}{2 (3\%)\xspace}
\newcommand{\othervalidation}{10 (16\%)\xspace}
\newcommand{\totalvalidation}{61 (100\%)\xspace}

\newcommand{\headerlenghtvaluedistribution}{8 (13\%)\xspace}
\newcommand{\optionlenghtvaluedistribution}{8 (13\%)\xspace}
\newcommand{\headerfieldvaluedistribution}{24 (39\%)\xspace}
\newcommand{\optionvaluedistribution}{2 (3\%)\xspace}
\newcommand{\truncatedpacketdistribution}{15 (25\%)\xspace}
\newcommand{\otherdistribution}{4 (7\%)\xspace}
\newcommand{\totaldistribution}{61 (100\%)\xspace}

\newcommand{\statelessstatefulness}{43 (70\%)\xspace}
\newcommand{\requirestcpstatefullness}{6 (10\%)\xspace}
\newcommand{\requiresrplstatefullness}{1 (2\%\xspace}
\newcommand{\requiresblestatefullness}{2 (3\%)\xspace}
\newcommand{\requiresmqttstatefullness}{4 (7\%)\xspace}
\newcommand{\requiressixlowpanstatefullness}{3 (5\%)\xspace}
\newcommand{\requireslinklayerstatefullness}{2 (3\%)\xspace}
\newcommand{\totalstatefullness}{61 (100\%)\xspace}

\newcommand{\toolname}{EmNetTest\xspace}
\newcommand{\toolnames}{EmNetTest's\xspace}
\newcommand{\tester}{Packetdrill++\xspace}
\newcommand{\ensbench}{ENSBench\xspace}

\begin{document}

\newcommand{\MyTitle}[1]{}
\renewcommand{\MyTitle}{Characterizing and Systematically Detecting Vulnerabilities in Embedded Network Stacks}
\renewcommand{\MyTitle}{Before you fuzz: A Systematic Testing Approach to Detecting Vulnerabilities in Embedded Network Stacks}
\renewcommand{\MyTitle}{Test Before You Fuzz: An Evaluation of Systematic Testing in Embedded Network Stacks}
\renewcommand{\MyTitle}{A Systematic Testing Framework for Detecting Packet Validation Vulnerabilities in Embedded Network Stacks}
\renewcommand{\MyTitle}{Systematically Testing Packet Validation in Embedded Network Stacks}
\renewcommand{\MyTitle}{EmNetTest: Systematically Detecting Packet Processing Vulnerabilities in Embedded Network Stacks}
\renewcommand{\MyTitle}{Systematically Detecting Packet Validation Vulnerabilities in Embedded Network Stacks}

\title{\MyTitle}

\ifANONYMOUS
\author{\IEEEauthorblockN{Anonymous Author(s)}}
\else
\author{\IEEEauthorblockN{Paschal C. Amusuo}
\IEEEauthorblockA{\emph{Electrical and Computer Engineering} \\
\emph{Purdue University}\\
West Lafayette, USA \\
pamusuo@purdue.edu}

\and
\IEEEauthorblockN{Ricardo Andr\'{e}s Calvo M\'{e}ndez}
\IEEEauthorblockA{\emph{Systems and Computer Engineering} \\
\emph{Universidad Nacional de Colombia}\\
Bogotá, Colombia\\
rcalvom@unal.edu.co}

\and
\IEEEauthorblockN{Zhongwei Xu}
\IEEEauthorblockA{\emph{Systems and Computer Engineering} \\
\emph{Xi'an JiaoTong University}\\
Xi'an Shaanxi, China \\
2206515211@stu.xjtu.edu.cn}

\linebreakand       
\IEEEauthorblockN{Aravind Machiry}
\IEEEauthorblockA{\emph{Electrical and Computer Engineering} \\
\emph{Purdue University}\\
West Lafayette, USA \\
amachiry@purdue.edu}

\and
\IEEEauthorblockN{James C. Davis}
\IEEEauthorblockA{\emph{Electrical and Computer Engineering} \\
\emph{Purdue University}\\
West Lafayette, USA \\
davisjam@purdue.edu}
}
\fi

\maketitle
\thispagestyle{plain}

\begin{abstract}

Embedded Network Stacks (ENS) enable low-resource devices to communicate with the outside world, facilitating the development of the Internet of Things and Cyber-Physical Systems. 
Some defects in ENS are thus high-severity cybersecurity vulnerabilities: they are remotely triggerable and can impact the physical world.
While prior research has shed light on the characteristics of defects in many classes of software systems,
  no study has described the properties of ENS defects nor identified a systematic technique to expose them.
The most common automated approach to detecting ENS defects is feedback-driven randomized dynamic analysis (``fuzzing''), a costly and unpredictable technique.

This paper provides the first systematic characterization of cybersecurity vulnerabilities in ENS.
We analyzed \vulnerabilitycount vulnerabilities across \networkstackstudiedcount open-source ENS.
Most of these ENS defects are 
  concentrated in the transport and network layers of the network stack,
  require reaching different states in the network protocol,
  and
  can be triggered by only 1-2 modifications to a single packet.
We therefore propose a novel systematic testing framework that
  focuses on the transport and network layers,
  uses seeds that cover a network protocol's states,
  and
  systematically modifies packet fields.
We evaluate this framework on \networkstackcount ENS and replicated \recreatedvulncount of the \possiblerecreationcount reported IP/TCP/UDP vulnerabilities.
On recent versions of these ENSs, it discovered \newvulncount novel defects (6 assigned CVES) during a bounded systematic test that covered all protocol states and made up to 3 modifications per packet. 
We found defects in 3 of the 4 ENS we tested that had not been found by prior fuzzing research. 
Our results suggest that fuzzing should be deferred until after systematic testing is employed. 
\end{abstract}


\section{Introduction}

\emph{\acfp{ENS}} are software components that enable network communication on embedded systems.
There are several~\acp{ENS} with varied architectures tailored to the semantics of individual embedded operating systems, such as Contiki-ng~\cite{stanoev_contiki-ng_nodate} and FreeRTOS~\cite{freertos_freertos_nodate}.
Unlike the network stacks used by regular operating systems, \acp{ENS} run on embedded systems with limited or no vulnerability protections~\cite{yu_building_2022}.
As a result, vulnerabilities in~\acp{ENS} are severe and could be remotely exploitable.
In the last five years, many critical defects have been discovered and reported in these \acp{ENS}~\cite{newman_operating_nodate, noauthor_amnesia33_nodate, forescout_project_nodate}.
Detecting cybersecurity vulnerabilities in~\acp{ENS} remains an important challenge for securing the Internet of Things.

Automated software testing techniques for network stacks use formal methods, static analysis, and dynamic analysis to detect vulnerabilities.
Formal methods, \eg model checking, provide strong guarantees~\cite{lockefeer_formal_2016, smith_formal_nodate, musuvathi_model_nodate, zaostrovnykh_verifying_2019, zhang_automated_nodate, pirelli_automated_2022} but are costly to apply and maintain.
Static analyses efficiently find defects~\cite{chen_static_2015, zhang_statically_2021, kothari_finding_2011, chang_inputs_2009, pedrosa_analyzing_2015}, but must be tuned to defect patterns and generate false positives.
Dynamic analysis is promising, especially fuzzing~\cite{pham_aflnet_2020, natella_stateafl_2022, andronidis_snapfuzz_2022, zou_tcp-fuzz_2021, poncelet_so_2022}. 
But while fuzzing ensures no false positives, it offers limited guarantees. 
No dynamic works examine the systematic testing of~\acp{ENS} and consequently provide guarantees.

Our goal was to develop a systematic dynamic testing technique, one that could provide certain guarantees about the security of the ~\ac{ENS} under test.
But what guarantees should be prioritized?
Several studies~\cite{anandayuvaraj2022FailureAwareSDLC, norman1990commentary, johnson_software_2000} show that defect patterns recur in software.
Thus, identifying the characteristic defects can help prevent such defects in the future. 
We analyzed~\vulnerabilitycount security defects that were previously reported across~\rtoscount embedded network stacks to understand defect patterns.
We found that most~\acp{ENS} vulnerabilities occur because an~\ac{ENS} directly used certain fields in packet headers without proper validation.
Invalid values of such fields lead to out-of-bound (OOB) reads, buffer overflows, and integer wraparounds. 
We call these \textbf{packet validation vulnerabilities}.
Our study also revealed that the test suites used in~\acp{ENS} are inadequate to detect this recurring class of defect.
To dynamically detect packet validation vulnerabilities, an approach must be
  systematic in varying fields (many different packet fields were problematic),
  able to reach different protocol states (many different protocol states were problematic),
  and
  able to find memory errors (most vulnerabilities involve OOB memory access).

Based on this analysis, we propose~\toolname, an automated and systematic framework for dynamic testing of~\acp{ENS}. \toolname possesses three characteristics that enable it to uncover known vulnerability patterns in \acp{ENS}. 
\emph{(1) Systematic packet generation:} \toolname systematically generates validly constructed packets with invalid header fields or truncated headers. 
\emph{(2) Stateful:} \toolname provides sequences of packets that get the ENS to different protocol states before packet injection. \emph{(3) Memory focused:} \toolname uses address sanitizers with dynamic memory poisoning to detect all memory corruptions. We implemented \toolname using~\packetdrill which provides necessary scripting support for testing network stacks. We enhanced~\packetdrill to support mutating arbitrary network packets.

\begin{figure*}[h]
\begin{center}

\includegraphics[width=0.8\textwidth]{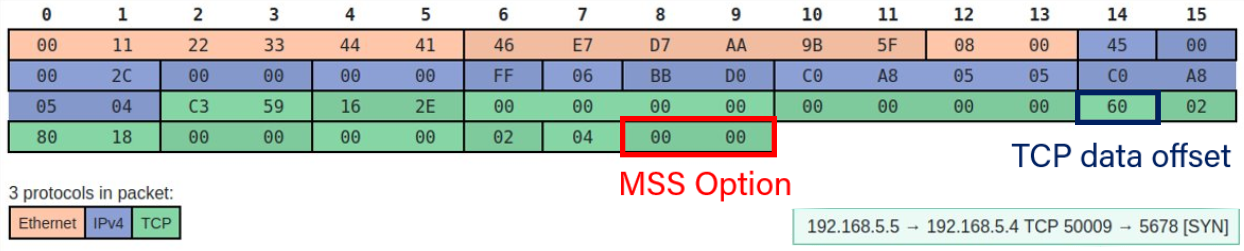}

\captionof{figure}{
    A hex representation of a TCP packet showing the TCP header length field and the TCP MSS Option Value.
    \cref{listing:NFA-Algorithm} describes two CVEs associated with these fields in the FreeRTOS \ac{ENS}.
 }
 \label{fig:tcp-vuln-packet}

\end{center}
\end{figure*}

We evaluated \toolname on 4 of the 6 \acp{ENS} whose vulnerabilities we studied: FreeRTOS, Contiki-ng, lwIP, and PicoTCP. 
We also created \ensbench, a dataset of~\recreatedvulncount vulnerabilities by re-introducing previously known vulnerabilities into recent versions of the ~\acp{ENS}.
Our evaluation showed that~\toolname replicated all the~\recreatedvulncount vulnerabilities we attempted.
In addition,~\toolname found~\newvulncount new vulnerabilities (zero days), which can be remotely exploited by any user and potentially allow arbitrary code execution.
We compared our framework with fuzzing. We ran 4 fuzzers from the Poncelet~\etal benchmarks \cite{poncelet_so_2022} on the latest version of Contiki-ng (which contains 5 vulnerabilities) and found that within 24 hours, no fuzzer detected any of the vulnerabilities.

Our work shows the importance and effectiveness of systematic testing for detecting critical software defects.
We invite the community to explore systematic testing approaches, beyond the current trend of automated randomized testing (fuzzing).

In summary, we contribute:
\begin{enumerate}[leftmargin=*]
\item We perform the first comprehensive study (\cref{sec:vulstudy}) of~\vulnerabilitycount reported ENS vulnerabilities, understand their root causes, and provide insights into the packet sequences that trigger these vulnerabilities.
\item We designed and implemented~\toolname (\cref{sec:emnetstest}), an automated systematic testing framework for~\ac{ENS}. Our evaluation shows that~\toolname{} effectively finds known and new vulnerabilities in~\ac{ENS}.
\item As part of our framework, we implemented~\pd, an extended version of~\packetdrill that facilitates adversarial testing of network stacks and can be used independently of our testing framework.
\item\textbf{\ourdataset{}}: A dataset of~\recreatedvulncount recreated and~\newvulncount new vulnerabilities, packaged into recent versions of~\acp{ENS} to support the evaluation of other defect detection tools. \toolname detects all vulnerabilities in this dataset.
\end{enumerate}

\noindent
Our vulnerability analysis and the implementation of \toolname are available (\cref{sec:DataAvailability}).

\definecolor{field1}{RGB}{74, 144, 226}
\definecolor{field2}{RGB}{65,117,5}

\setlength{\fboxrule}{1pt} 

\begin{listing}
  \centering
  \caption{
   CVE-2018-16523 and CVE-2018-16524: Snippet showing a divide-by-zero defect triggered by the TCP MSS Option (green) and an out-of-bound read (blue) triggered by the TCP Data Offset.
   Both are in the FreeRTOS network stack.
   \toolname can recreate these vulnerabilities (\cref{tab:recreated-vulns}).
  }
  \label{listing:NFA-Algorithm}
  \begin{tcolorbox} [width=\linewidth, colback=white!30!white, top=1pt, 
  bottom=1pt, left=2pt, right=2pt]
\begin{minted}[
    fontsize=\footnotesize,
    linenos,
    gobble=2, % Remove unnecessary indentation -- the line numbers make this clear enough
    xleftmargin=0.5cm, % Otherwise we start in the left margin...
    escapeinside=||
]{c}
static void prvCheckOptions(...) {
 const unsigned char *pucPtr = ... ;
 |\fcolorbox{field1}{white}{const unsigned char *pucLast = pucPtr +}|
 |\fcolorbox{field1}{white}{(((pxTCPHeader->ucTCPOffset >> 4) - 5) << 2);}|
 while(pucPtr < pucLast){
  ...
  else if((|\fcolorbox{field1}{white}{pucPtr[0] \textcolor{red}{\faBug}}| == TCP_OPT_MSS) && 
    (pucPtr[1] == TCP_OPT_MSS_LEN)) {
   |\fcolorbox{field2}{white}{uxNewMSS = usChar2u16(pucPtr + 2);}|
   if(pxSocket->u.xTCP.usInitMSS > uxNewMSS){
    ...
    pxTCPWindow->xSize.
    |\fcolorbox{field2}{white}{ulRxWindowLength = ((uint32\_t) uxNewMSS) *}| 
    |\fcolorbox{field2}{white}{(pxTCPWindow->xSize.ulRxWindowLength / }|
    |\fcolorbox{field2}{white}{((uint32\_t) uxNewMSS \textcolor{red}{\faBug}));}|
       ...
  }}
  pucPtr += ...
  ...
}}

  \end{minted}
\end{tcolorbox}
\end{listing}

\section{Background}
\label{sec:background}

\subsection{Embedded Network Stacks (ENS)}
\acfp{ENS} enable network connectivity for embedded systems.
\acp{ENS} are either part of an embedded operating system (Integrated ENS)~\cite{freertos_freertos_nodate, oikonomou_contiki-ng_2022} or stand-alone libraries (Standalone ENS)~\cite{lwip_lwip_nodate, picotcp_picotcp_nodate}.
\acp{ENS} follow a layered software architecture, each layer implementing a specific protocol on the TCP/IP stack.

\acp{ENS} vulnerabilities pose a greater threat than those of regular network stacks due to the absence of operating systems and hardware vulnerability protection mechanisms.
Regular operating systems, \eg Linux, provide more protection in their OS design.
This includes features such as Address Space Layout Randomization (ASLR), Data Execution Prevention (DEP), address space isolation, and Stack Canaries~\cite{cowan1998stackguard} that prevent the exploitation of memory vulnerabilities.
Also, modern processors include no-execute (Nx) regions that prevent the unauthorized execution of codes in sensitive memory regions~\cite{noauthor_defeating_nodate}.
Many embedded OSes and processors lack these features~\cite{yu_building_2022,abbasi_challenges_2019}, increasing the ease of vulnerability exploitation.

\acp{ENS} are designed for embedded systems, which are resource-constrained, have real-time requirements, and often lack common library support.
\acp{ENS} are also tailored to the underlying embedded operating system's threading and scheduling semantics.
Consequently, they differ from regular operating systems' network stacks, which use the POSIX standard~\cite{ieee_ieee_2001} for portability.
For example, the~\inlinecode{accept} call in FreeRTOS blocks until a successful TCP connection is established or the timeout elapses. 
The~\inlinecode{accept} call in lwIP is non-blocking and defines a callback that would be called on successful connection establishment.
Meanwhile, Contiki-ng has no~\inlinecode{accept} syscall. Instead, it uses an event-driven callback for all events, including a Socket connection event.

\subsection{Internet Protocols and Network Packets}
\begin{figure}[h]
 \centering
 \includegraphics[width=0.95\columnwidth]{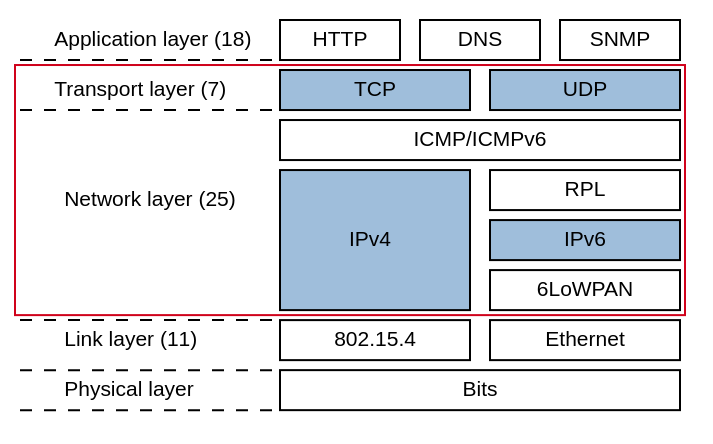}
 \captionof{figure}{
   Layers of the TCP/IP stack.
   Our tool targets layers in the red box (specific protocols in blue).
   Values in parentheses indicate number of analyzed CVEs in each layer (\cref{subsubsec:vuln-layers}).
   }
 \label{fig:ip-stack}
\end{figure}

In this work, we focus on Internet Protocol (IP) or TCP/IP suite, which includes various protocols that specify how data should be packaged, addressed, and routed~\cite{robert_t_braden_requirements_1989, braden_requirements_1989}.
The TCP/IP suite is organized into a layered architecture (\cref{fig:ip-stack}).
An Internet packet has elements for each layer of this architecture, recursively structured as headers associated with one layer and a payload associated with the next (\cref{fig:tcp-vuln-packet}).
The protocol's implementation processes the corresponding headers at each level and passes the payload along.

\subsection{\ac{ENS} Vulnerabilities}

\acp{ENS} are usually implemented in C/C++ for performance and compatibility reasons.
They handle complex packet structures across multiple layers.
Hence, \acp{ENS} are prone to defects that can be cybersecurity vulnerabilities. 
With the absence of sufficient protection mechanisms, the vulnerabilities in an~\ac{ENS} can be exploited to either disable or remotely control the entire system.
Furthermore, these vulnerabilities can be triggered remotely by any use with network access to the system.

\cref{listing:NFA-Algorithm} shows the snippet corresponding to two vulnerabilities, \mbox{CVE-2018-16523} and \mbox{CVE-2018-16524}.
\mbox{CVE-2018-16523} (divide by zero) occurs because the TCP MSS option value,~\inlinecode{uxNewMSS}, is used as the divisor to calculate~\inlinecode{RxWindowLength}. A TCP packet with an MSS value of zero will lead to a divide-by-zero error.
\mbox{CVE-2018-16524} (out-of-bounds read) occurs because offset in the header,~\ie~\inlinecode{ucTCPOffset} is used to compute a pointer address~\inlinecode{pucLast}, which is later read through~\inlinecode{pucPtr}.
These vulnerabilities are triggerable remotely without authorization by sending TCP packets.
Furthermore, these vulnerabilities can be exploited to gain control of the system because of the lack of isolation mechanisms in embedded systems.

\section{Related Work} \label{relatedWork}

\paragraph{Traditional Testing} Many \acp{ENS} incorporate test suites that help the maintainers validate the various functionalities they develop. 
As shown in~\cref{sec:vulstudy}, these test suites are inadequate. Although automated test generation tools~\cite{fraser_evosuite_2011, pacheco_feedback-directed_2007, beyer_advances_2022} exist, the tests generated by them are inadequate at finding faults~\cite{serra_effectiveness_2019}.
Furthermore, domain knowledge is required to use these automated test-generation tools effectively.

Research and commercial tools exist to facilitate the easy development of test suites for network stacks.
\packetdrill~\cite{cardwell_packetdrill_2013}, a network stack testing tool that enables the use of scripts to test the end-to-end correctness behavior of network stacks.
\packetdrill focuses on testing functionality and always generates valid packets,~\ie has valid and well-formed headers.
However, as found in~\cref{sec:vulstudy}, most vulnerabilities occur because of invalid values in packet headers.
InterWorking Labs has commercial testing solutions for testing network protocols and also uses malformed packets~\cite{interworkinglabs}.
However, access costs over \$10,000,\footnote{This quote was provided to us through personal communication.} limiting its adoption in open-source projects and the low-margin embedded systems marketplace~\cite{gopalakrishna_if_2022}.

\paragraph{Formal Methods} Several tools~\cite{zaostrovnykh_verifying_2019, zhang_automated_nodate, pirelli_automated_2022} have explored formal methods for verifying network functions.
Zaostrovnykh~\cite{zaostrovnykh_verifying_2019} and Pirelli~\cite{pirelli_automated_2022} developed formal verification tools to automatically prove that a network function conforms to a provided specification.
However, these techniques do not apply to multithreaded programs such as~\ac{ENS}.
Microsoft's Project Everest~\cite{microsoft_project_nodate} verifies various components of HTTPs and has provided verified implementations of some cryptographic libraries. 
FreeRTOS, maintained by AWS, also verifies their network stack implementation, FreeRTOS+TCP~\cite{Chong2021}.
These formal methods are used to verify specific correctness properties of the network protocol implementations and do not make complete guarantees about their security. 
As shown by Fonseca~\etal~\cite{fonseca_empirical_2017}, formal methods guarantees are only as good as their underlying assumptions.
Hence, we still need to assess the security of these systems through dynamic testing.

\paragraph{Fuzzing} Fuzzing~\cite{manes2019art} has found many software defects.
From a network perspective, fuzzing has been mainly explored to find bugs in network applications.
AFLNet \cite{pham_aflnet_2020}, StateAFL \cite{natella_stateafl_2022}, and SnapFuzz \cite{andronidis_snapfuzz_2022} are three recent works in this direction. 
These works focus on setting up a proper communication channel with a network application and sending test data to the application through well-formed network packets.
A recent work, TCPFuzz~\cite{zou_tcp-fuzz_2021} uses fuzzing and differential testing to detect semantic vulnerabilities in the transport layer.
TCPFuzz always generates valid packets and cannot find vulnerabilities arising from invalid packets.

Poncelet~\etal~\cite{poncelet_so_2022} applied several state-of-the-art fuzzing tools to test individual functions of the Contiki-ng~\ac{ENS}. 
They reported that testing lower-layer functions does not get deep penetration.
Conversely, directly testing upper network layers increases the rate of false positives as some inputs and corresponding packets are impossible as lower layers will reject them.
Furthermore, they fail to trigger code paths that require the network stack to be in a particular state. 
This is demonstrated in our evaluation (\cref{sec:evaluation}) where~\toolname found various vulnerabilities in the well-test portions of Contiki-ng.

\paragraph{Vulnerability Studies} Several researchers have studied vulnerabilities' characteristics in different software systems \cite{liu_large-scale_2020, jimenez_empirical_2016, cai_understanding_2019, al-boghdady_presence_2021, mcbride_security_2018}. 
Most of these works focus on well-provisioned systems, \eg desktop and web software. 
Few works study vulnerabilities in embedded systems.
Al-Boghdady~\etal \cite{al-boghdady_presence_2021} studied the characteristics of security vulnerabilities in IoT operating systems. While they focused on characterizing the CWEs (Common Weakness Enumeration) reported by static analysis tools, their study does not cover how these vulnerabilities are triggered or detected.  
Similar to our work, Malik \& Pastore examined CVEs in Edge frameworks and found that (1) the network components were a common source of CVEs, and (2) specific values were often problematic, but did not go into detail on \acp{ENS} nor evaluate a solution~\cite{malik2023empirical}.
Other industry practitioners have also published reports of security vulnerability analyses of \acp{ENS} they conducted.
For example,
  Zimperium \cite{zimperium_labs_freertos_nodate} published a blog post containing details of the vulnerabilities they discovered in FreeRTOS,
  and
  Forescout published a report containing an analysis of the 33 vulnerabilities they found and a list of observed common anti-patterns~\cite{noauthor_amnesia33_nodate}.
No prior work systematically analyzes vulnerabilities in ~\acp{ENS}.

\section{Knowledge Gaps and Research Questions}

This work aims to fill two gaps.
First, no study characterizes cybersecurity vulnerabilities in \acp{ENS}.
Second, no dynamic system exists to systematically detect ENS cybersecurity vulnerabilities.
We ask:

\iftrue



\begin{smalldescription}
\item[Theme 1:] \textbf{Vulnerability analysis}

\item[RQ1:] What are the types and root causes of vulnerabilities?
\item[RQ2:] What packet sequences trigger ENS vulnerabilities?  
\end{smalldescription}

\vspace{0.05cm}
\noindent


\begin{smalldescription}
\item[Theme 2:] \textbf{State of practice for packet validation testing}
\item[RQ3:] Are ENS tested for packet validation vulnerabilities?
\end{smalldescription}

\begin{smalldescription}
\item[Theme 3:] \textbf{Evaluating systematic testing with \toolname}
\item[RQ4:] To what extent can bounded systematic testing uncover packet validation vulnerabilities? 
\end{smalldescription}

\fi

\section{ENS Vulnerabilities and Testing (RQ1-3)}
\label{sec:vulstudy}

This section presents methodology and results for RQ1-3. 
To summarize our findings, ENS CVEs are typically \emph{packet validation vulnerabilities}.
The studied ENS incorrectly handle packets that are slightly malformed, sometimes from a particular protocol state.
In \lessthreemutationspercent\% of CVEs, 1-2 fields are incorrect. Repairs often involve a single \code{if}-statement.

\begin{table*}[ht]
\caption{
    Embedded network stacks (integrated and standalone) whose CVEs we examined.
    C/C++ LoC (source, not tests) measured with cloc\cite{cloc-tool}; for integrated ENS we measured only the network implementation.
    GitHub data as of May 2023.
}
\label{tab:selected-stacks}
\begin{center}
\begin{small}
\begin{tabular}{ccccc|cc}
\toprule
\textbf{Name} & \textbf{Size (LOC)} & \textbf{GitHub stars} & \textbf{GitHub forks} & \textbf{\# CVEs studied} & \textbf{\# CVEs recreated} & \textbf{\# new vulns.} \\
\toprule
FreeRTOS(+TCP) & \freertossize & \freertosstars & \freertosforks & \freertosvulncount & \freertosrecreatedvuln/\freertosnettransvuln & \freertoscvediscovered \\
Contiki-ng  & \contikisize & \contikistars & \contikiforks & \contikivulncount & \contikirecreatedvuln/\contikinettransvuln & \contikinewvuln \\
Zephyr  & \zephyrsize & \zephyrstars & \zephyrforks & \zephyrvulncount & \zephyrcverecreated & \zephyrcvediscovered \\
\midrule
PicoTCP     & \picotcpsize & \picotcpstars & \picotcpforks & \picotcpvulncount & \picotcprecreatedvuln/\picotcpnettransvuln & \picotcpnewvuln \\
LwIP        & \lwipsize & \lwipstars & \lwipforks & \lwipvulncount & \lwipcverecreated & \lwipnewvuln \\
FNET        & \fnetsize & \fnetstars & \fnetforks & \fnetvulncount & \fnetcverecreated & \fnetcvediscovered \\
\bottomrule
\end{tabular}
\end{small}
\end{center}
\end{table*}

\subsection{Methodology}

\subsubsection{Repository Selection}
We studied both integrated ENS and standalone ENS (\cref{tab:selected-stacks}).
We selected~\acp{ENS} integrated into major open-source embedded operating systems.
From lists in survey papers~\cite{hahm_operating_2016, silva_operating_2019},
we selected three embedded OSes with over 1K GitHub stars: 
  Zephyr (maintained by Linux Foundation),
  Contiki-ng (Supported by Swedish Research Institute),
  and 
  FreeRTOS (maintained by AWS).
From a previous vulnerability study~\cite{forescout_project_nodate}, we selected the top-3 actively-maintained repositories (by GitHub stars) with reported CVEs. 
These were PicoTCP, LwIP, and FNet. 

\subsubsection{Data Collection}
We obtained vulnerability reports (CVEs) from the National Vulnerability Database (NVD)~\cite{nvd}.
We searched the NVD for the associated project.
For integrated~\acp{ENS}, we only considered vulnerabilities in the networking stack.
There were \totalvulncount total CVEs. 
We discarded \nontechnicalvulncount CVEs that omitted technical vulnerability details. After preliminary analysis, we observed \vulnerabilitycount of the remaining 66 vulnerabilities were caused by the poor validation  of packets received by ENS. We termed these \emph{packet validation (PV) vulnerabilities}. We removed the \nonpvvulncount non-PV vulnerabilities. 

\subsubsection{Data Analysis}
One author analyzed each vulnerability report and technical details, including
  screenshots explaining vulnerable code,
  links to the vulnerability's GitHub issue,
  and
  the repairing pull request (PR).\footnote{During this analysis, we found 3 new CVEs (excluded from our analysis).}
  We indicate the specific extracted features below --- these are a typical set of features in software failure analysis~\cite{amusuo2022SoftwareFailureAnalysis}.
For soundness, a second author 
analyzed a random sample of \multiratedlabels vulnerabilities.
We measured interrater agreement using Cohen's Kappa score~\cite{cohen1960coefficient}.
We obtained $\kappa$=\cohenkappa, indicating substantial agreement~\cite{landis1977application}.

\subsection{RQ1: Vulnerability Characteristics} \label{rq1}

\begin{tcolorbox} [width=\linewidth, colback=yellow!30!white, top=1pt, bottom=1pt, left=2pt, right=2pt]
\textbf{Finding 1}: 
Memory Out-of-Bound Read and Write are the most common vulnerabilities (\memoryvulnpercent\%).

\textbf{Finding 2:}
Missing length field validation and Missing packet size validation are the most frequent root causes and account for \lengthandpacketsizepercent\% of vulnerabilities.

\textbf{Finding 3:}
The network layer contains most vulnerabilities (\netvulnpercent\%), followed by the application layer (\appvulnpercent\%). Vulnerabilities are also found in every layer of the stack. 
\end{tcolorbox}


\noindent
We describe CVE types, root causes, and affected components. 

\subsubsection{Vulnerability Types}
\label{subsubsec:vuln-types}
First, we group CVEs according to their Common Weakness Enumeration (CWE)~\cite{cwe}.
%
\begin{table}
\caption{
    Proportion of CVE types.
    ``Others'': double-free, DNS cache poisoning, division-by-zero, and infinite loops.
}
\label{tab:vulnerability-types}
\begin{center}
\begin{small}
\begin{tabular}{lr}
\toprule
\textbf{Type} & \textbf{\# CVEs (\%)} \\
\toprule
Out-of-Bounds Read (CWE 125,126,200) & \outofboundsreadtype\\
Out-of-Bounds Write (CWE 120,121,122,787) & \outofboundswritetype \\
Integer Overflow (CWE 191) & \integeroverflowtype \\
Integer Underflow (CWE 190) & \integerunderflowtype\\
Null-pointer dereference (CWE 476) & \nullpointerdereferencetype \\
Other & \othertype \\
\midrule
Total & \totaltype \\
\bottomrule
\end{tabular}
\end{small}
\end{center}
\end{table}
\cref{tab:vulnerability-types} shows the result by this taxonomy.
Memory over-read/write (the first two rows) comprise \memoryvulnpercent\% of the vulnerabilities.

\subsubsection{Implementation Root Causes}
\label{subsubsec:vuln-patterns}
We studied code and repairs to learn the implementation-level root causes of CVEs. 
\cref{tab:vulnerability-patterns} groups these into several recurring patterns.
Roughly \lengthandpacketsizepercent\% of CVEs in~\acp{ENS} (first two rows) result from missing checks on length fields and data packet size.
Two examples:

\begin{table}
\caption{
    Implementation-level root causes of CVEs.
}
\label{tab:vulnerability-patterns}
\begin{center}
{
\small
\begin{tabular}{lr}
\toprule
\textbf{Root cause} & \textbf{\# CVEs (\%)} \\
\toprule
Missing length field validation & \missinglenghtfieldvalidation \\
Missing packet size validation & \missingpacketsizevalidation \\
Missing header value validation & \missingheadervaluevalidation \\
Missing integer wraparound validation & \missingintegerwraparoundvalidation \\
Other & \othervalidation \\
\midrule
Total & \totalvalidation \\
\bottomrule
\end{tabular}
}
\end{center}
\end{table}

\begin{itemize}
    \item \textbf{Missing length field validation (CVE-2018-16524):}
    FreeRTOS uses the TCP header length field to calculate the size of the TCP options region.
    However, it fails to validate the length value. Consequently, an invalid length value results in arbitrary memory read.

    \item \textbf{Missing packet size validation (CVE-2022-36054):}
    Contiki-ng receives a 6LoWPAN packet and after header compression, copies the packet into a buffer. If the packet is the first fragment of a fragmented packet, only 148 bytes are allocated. Contiki-ng doesn't verify the received packet size before copying it into this buffer. Consequently, a buffer overflow could result in a remote code execution attack.

\end{itemize}

\subsubsection{Vulnerable Layers}
\label{subsubsec:vuln-layers}

The left column of~\cref{fig:ip-stack} shows the distribution of CVEs across the ENS layers. 
%
The top layers for CVEs are network (\netvulnpercent\%) and application (\appvulnpercent\%). 

\subsection{RQ2: Packet Sequences That Trigger CVEs} \label{rq2}

\begin{tcolorbox} [width=\linewidth, colback=yellow!30!white, top=1pt, bottom=1pt, left=2pt, right=2pt]
\textbf{Finding 4}: \lessthreemutationspercent\% of vulnerabilities depend on one or two fields and consequently can be triggered with a maximum of two field changes. \numdependentfields different fields contribute to these vulnerabilities.

\textbf{Finding 5}: \statefulvulnpercent\% of CVEs are stateful, \eg involving an existing connection or a specific protocol state.
\end{tcolorbox}

Here, we study packet sequences that can trigger these CVEs.
Each packet sequence has a prefix (\ie state prefix) $p_1 p_2 \ldots p_{k-1}$ that brings the ENS to a vulnerable state, followed by the vulnerability-triggering packet $p_k$.
For instance, consider a vulnerability in processing a TCP FIN packet.
To trigger the vulnerability, we first need to send packets that can set the~\ac{ENS} to a state where it accepts a FIN packet.
Then, we send a FIN packet triggering the vulnerability. 
Understanding both parts enables a testing scheme to uncover real CVEs.



\subsubsection{Properties of the vulnerability-triggering packets ($p_{k}$)}
\label{subsubsec:vuln-fields}

Here, we investigate two aspects: 
(1) \emph{Root Cause Fields} ($RC_{f}$): Which incorrectly-handled fields result in vulnerabilities? 
(2) \emph{Dependent Fields} ($D_{f}$): How many fields of a packet does a vulnerability depend on?

For instance, consider \mbox{CVE-2018-16599}, which is caused by the incorrect validation of the UDP header length field. The vulnerability can be triggered only if the NBNS Type field  is NET\_BIOS (0x0020) and the NBNS Flags field indicates a response packet (0x8000).
Here, $RC_{f}$ = 1 (for the length field), whereas $D_{f}$ = 3 (for the length, type, and flags fields).

\cref{tab:vulnerable-fields} shows $RC_{f}$ and the number of CVEs resulting from it.
We see that 57 (93\%) of CVEs arise from fields in the protocol headers and options that are incorrectly handled.
\cref{tab:field-dependencies} shows the distribution of CVEs according to $D_{f}$.
Most vulnerabilities (58, or 95\%) have $D_{f} \leq 2$.
However, it is not just one field that is problematic --- \numdependentfields different fields across \numprotocols protocols contribute to the \vulnerabilitycount CVEs.

%
%

\ifJOURNAL
\else
\begin{table}
\caption{
    Distribution of CVEs based on the incorrect fields ($RC_{f}$) in the CVE-triggering packet.
    These fields often included those specifying the length of the packet or option component (rows 1-2),
    or specific values of other fields or options (rows 3-4).
    Often, the packet was truncated (row 5).
}
\label{tab:vulnerable-fields}
\begin{center}
{
\small
\begin{tabular}{lc}
\toprule
\textbf{Type} & \textbf{Count(\%)} \\
\toprule
Header length value & \headerlenghtvaluedistribution \\
Option length value & \optionlenghtvaluedistribution \\
\midrule
Header field value & \headerfieldvaluedistribution \\
Option value &\optionvaluedistribution \\
\midrule
Truncated packet & \truncatedpacketdistribution \\
\midrule
Others & \otherdistribution \\
\midrule
Total & \totaldistribution \\
\bottomrule
\end{tabular}
}
\end{center}
\end{table}
\fi

\begin{table}
\caption{
    Distribution of CVEs by \# of dependent fields ($D_{f}$).
}
\label{tab:field-dependencies}
\begin{center}
{
\small
\begin{tabular}{ll}
\toprule
\textbf{\# Dependent Fields} & \textbf{\# CVEs} \\
\toprule
1 & \onechange (\onechangeperc) \\
2  & \twochanges (\twochangesperc) \\
$>$ 2 & \moretwochanges (\moretwochangesperc) \\
\midrule
Total & \vulnerabilitycount (\vulnerabilitycountperc) \\
\bottomrule
\end{tabular}
}
\end{center}
\end{table}

\subsubsection{Properties of the packet sequence prefix}
\label{subsubsec:Vuln-SysConditions}


We studied the vulnerable code and execution path to identify any states involved. 
\cref{tab:protocol-states} shows that \statelessvulnpercent\% CVEs are stateless (can be triggered with a single packet/no prefix) and that the remaining \statefulvulnpercent\% (12 CVEs) depend on the state of the system.
Of these,
  13 CVEs, occurring on stateful protocols, require the protocol to be in a specific set of states.
  5 other CVEs depend on the properties of the previously-sent packet(s).

\begin{table}
\caption{
    Distribution of vulnerabilities based on the statefulness required to expose the vulnerability.
}
\label{tab:protocol-states}
\begin{center}
{
\small
\begin{tabular}{llc}
\toprule
\textbf{State Required} & \textbf{Protocol} & \textbf{\# CVEs} \\
\toprule
Stateless & -- & \statelessstatefulness \\
\midrule
Requires protocol state & TCP & \requirestcpstatefullness \\
Requires protocol state & RPL & \requiresrplstatefullness \\
Requires protocol state & BLE & \requiresblestatefullness \\
Requires protocol state & MQTT & \requiresmqttstatefullness \\
Requires packet sequence & 6LoWPAN & \requiressixlowpanstatefullness \\
Requires packet sequence & 802.15.4 & \requireslinklayerstatefullness \\
\midrule
Total & All & \totalstatefullness \\
\bottomrule
\end{tabular}
}
\end{center}
\end{table}

\subsection{RQ3: Testing Suite Characteristic}
\label{subsec:test-suite-characteristics}



We analyzed the test suites of four~\acp{ENS} to understand why the known CVEs, which we discussed in \cref{rq1}, existed. 
Based on the CVE characteristics, we looked for four aspects of validation:
  (1) Unit tests involving input packet processing operations with malformed input;
  (2) Capability of injecting specific (and possibly malformed) packets;
  (3) Tests involving packets with invalid headers (cf. \cref{tab:vulnerable-fields});
  and
  (4) Tests involving statefulness (cf. \cref{tab:protocol-states}).

\begin{tcolorbox} [width=\linewidth, colback=yellow!30!white, top=1pt, bottom=1pt, left=2pt, right=2pt]
\textbf{Finding 6}:~\acp{ENS} are validated using end-to-end simulation tests and unit tests. The actual implementations of these tests are unique in each ENS (no standard test framework).

\textbf{Finding 7:} While some~\acp{ENS} include packet injection tests, these are regression tests for specific CVEs. One ENS provides packet seeds and a harness for stateful fuzzing.

\textbf{Finding 8:} None of the~\acp{ENS} systematically check invalid header or option fields, nor include unit tests for the various operations performed on an input packet.
 
\end{tcolorbox}

\myparagraph{FreeRTOS}
FreeRTOS validates its ENS with end-to-end tests, unit tests, and formal verification. 
The end-to-end tests use the sockets interface to establish network connections and validate behaviors of the network stack. 
They provide (incomplete) memory safety proofs for main packet processing functions.
Not all functions are verified and the provided proofs depend on the corrections of some unverified functions.
FreeRTOS has some packet injection tests with invalid headers, but all cases are regressions for past CVEs. 

\myparagraph{Contiki-ng}
Contiki-ng is validated with network simulation using cooja~\cite{contiki-cooja-wiki}, a packet injection test, and fuzzing.
The network simulation tests involve various end-to-end tests under different simulated network environments.
Their packet injection tests use a fixed set of network packets.
These are mostly regression tests to check for previous defects or vulnerabilities. Their packet injection framework is also used for fuzzing.

\myparagraph{PicoTCP}
PicoTCP validates with unit tests and end-to-end demo applications.
The provided unit tests are mostly on non-packet related tasks, such as IP address-to-string conversion and socket tests.
The demo applications test supported protocols in different network environments.

\myparagraph{LwIP}
LwIP validates with unit tests, network stress testing, and fuzzing.
Their unit tests mostly test the output operations of the network stack, not the input packet processing functions.
Several unit tests configure the test socket to specific protocol states.
For stress, they measure the reliability of simulated networks while increasing the number of nodes and messages in the network.
They also provide a fuzzing harness and fuzzing seeds for the different protocols.

\section{\toolname: Design and Implementation}
\label{sec:emnetstest}
\subsection{Design Requirements}
\label{subsec:design-req}

Based on our findings from Theme 1, an automated testing framework to detect PV vulnerabilities in \acp{ENS} should have three characteristics:

\begin{itemize}
    \item \textbf{Ability to Detect Memory Issues:}
      Based on Finding 1, it should detect memory corruption vulnerabilities.
    \item \textbf{Systematic Packet Generation:}
      Based on Findings 2 and 4, it should systematically generate valid test packets with incorrect header values and truncated headers.
    \item \textbf{Stateful:}
      Based on Findings 3 and 5, it should drive the~\ac{ENS} stack to different protocol states for multiple protocols. 
\end{itemize}

\noindent
Such a framework would improve the state of the art in \ac{ENS} testing (cf.~\cref{relatedWork} and Findings 6-8).

\subsection{Design}

\begin{figure*}[h]
\begin{center}

\includegraphics[width=0.8\textwidth]{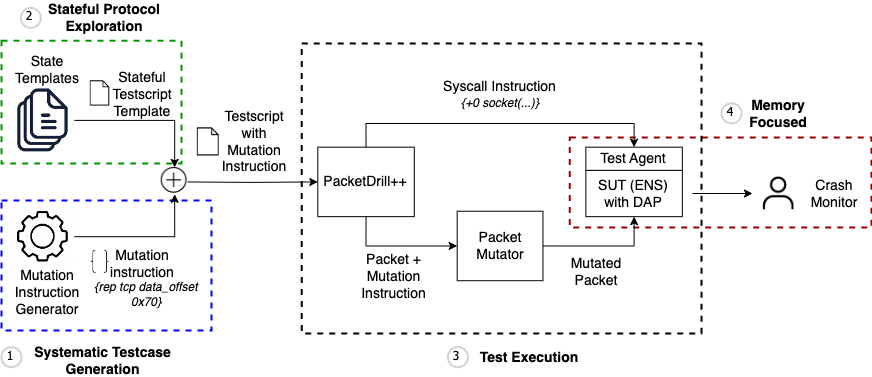}

\captionof{figure}{
   Overview of the design of \toolname.
   It systematically generates mutation instructions by repeatedly taking a protocol header, selecting combinations of fields, and iterating through possible values (blue box~\ding{172}).
   It achieves statefulness by using a set of test script templates that can explore different protocol states (green box~\ding{173}).
   Packetdrill++ interprets each test script and sends the syscall and packets to the SUT (ENS). 
   The per-ENS Test Agent maps received POSIX syscall instructions to appropriate behavior and execute the behavior  on the ENS (black box~\ding{174}). 
   The SUT is instrumented with dynamic poisoning and compiled with ASAN to aid the detection of memory corruption (red box~\ding{175}). 
}
\label{fig:fig-design-overview}

\end{center}
\end{figure*}

\noindent
\toolname meets these requirements.
\cref{fig:fig-design-overview} illustrates. 

\begin{itemize}
  \item \textbf{Memory focused:}
  Address Sanitization (ASAN)~\cite{serebryany2012addresssanitizer} with~\ac{ENS} specific instrumentation.
  \item \textbf{Systematic packet generation:}
    An ordered generation algorithm can systematically generate all packets but is prioritized for packets that trigger known PV CVEs. 
  \item \textbf{Stateful:}
    We build on the~\packetdrill tool~\cite{cardwell_packetdrill_2013}.
    It provides packet sequences that cover some relevant protocols.
    We extend it to additional protocols and employ a seed set of state-covering sequences (packet sequence prefixes).
\end{itemize}

\noindent

\definecolor{field1}{RGB}{74, 144, 226}
\definecolor{field2}{RGB}{65,117,5}

\setlength{\fboxrule}{1pt} 

\begin{listing}[H]
  \centering
  \caption{
   Systematic packet generation.
   The caller imposes order by working from smaller to larger $N$ and varying $S$.
  }
  \label{algorithm:SystematicOrderedGeneration}
  \begin{tcolorbox} [width=\linewidth, colback=white!30!white, top=1pt, 
  bottom=1pt, left=1pt, right=2pt]
\begin{minted}[
    fontsize=\footnotesize,
    linenos,
    autogobble,
    xleftmargin=0.5cm, % Otherwise we start in the left margin...
    escapeinside=||,
    mathescape
]{python}
# Input: Num. entities N, stride S, packet $p_k$
# Output: Yields next packet for this config.

# SELECTION of N fields and options
|$\left\lbrace f_1, \cdots, o_1, \cdots \right\rbrace$| = nextPacketEntities(N) # Generator

# INTERPOLATION
|$vals_{f_1}$| = interpolate(|$f_1$|, S)
|$\hdots$|
|$vals_{o_1}$| = interpolate(|$o_1$|, S)
|$\hdots$|

# GENERATION (uses Python itertools)
for |$f_{1}$| in |$vals_{f_1}$|:
  |$\hdots$|
    for |$o_{1}$| in |$vals_{o_1}$|:
      |$\hdots$|
        |$p_k$|.modify(|$f_{1}$|, |$\cdots$|, |$o_{i}$|, |$\cdots$|)
          yield |$p_k$|
# Caller sends prefix + |$p_k$| and checks result
\end{minted}
\end{tcolorbox}
\end{listing}

\subsubsection{Custom Address Sanitization with Dynamic Address Poisoning (DAP)}
Memory corruptions (\ie out-of-bounds read and write) in embedded systems may not lead to program crashes~\cite{muench2018you} (\texttt{SEGSEGV}).
ASAN is a well-known technique to convert memory corruptions into program crashes.
However, ASAN assumes the target application is using standard memory allocation and deallocation functions (\eg~\code{malloc}/\code{free}) --- which is not the case with~\acp{ENS}, as they use custom allocators.
To handle this, we use ASAN's Dynamic Address Poisoning (DAP) support.
\JD{I uncommented the following lines. I think they improve clarity. Please review for truthfulness and then remove this comment if they are OK.}
For each~\ac{ENS}, we modified its custom allocators such that after every allocation, the corresponding memory chunk will be unpoisoned (\ie OK to use).
Similarly, we modify deallocator or release functions to poison the corresponding memory chunk (\ie Invalid to use).

We modify packer copying routines to detect out-of-bound memory accesses during packet processing.
Specifically, after a packet is received and copied into the buffer, we poison the rest of the allocated buffer that is not covered by the received packet.
Then, ASAN will detect any bytes read or written beyond the bounds of the allocated buffer.



\subsubsection{(Ordered) Systematic Packet Generation}
We systematically generate ordered test packets.
\emph{Systematic} means all packets are generated.
\emph{Ordered} means an ordering over the packets such that likely-useful packets are generated early.

We describe our approach, formalized in~\cref{algorithm:SystematicOrderedGeneration}.
We assume a prefix sequence of packets $p_1 p_2 \ldots p_{k-1}$ to reach a desired protocol state, followed by test packet $p_{k}$ (\cref{rq2}).
The algorithm generates all valuations of $p_{k}$.

\ul{Systematic:}
Packet $p_k$ is a sequence of bytes consisting of required header fields, optional fields, and a payload.
The payload was not the cause of ENS CVEs (\cref{sec:vulstudy}) so we exclude it.
Different subsets of the header and option fields are selected to modify (generator on line 5).
We obtain possible values for each following an interpolation sequence from minimum (\eg \code{0x00}) to maximum (\eg \code{0xff}) along a stride $S$ (line 7).
All combinations are explored (line 13).
This ensures we can detect vulnerabilities that either depend on the minima or maxima, or on a range of values as determined by the stride.
To generate all $p_k$, choose maximum $N$ and a Stride of 1.

\ul{Ordered:}
We order the generation of $p_{k}$ starting from 0 entities (the original $p_k$), then 1 entity, and so on, discarding repeating packets.
This order places the likely-to-be-useful packets early in the sequence --- per~\cref{tab:field-dependencies}, most CVEs depend on at most 2 fields (\ie $D_{f} \leq 2$).
This suggests that most of the vulnerabilities could be found with $N=2$.
During ENS validation, engineers may parameterize by bounding the maximum number of fields to select $N$.
They may trade exhaustiveness vs. cost via the search stride $S$.

\ul{Handling truncate:}
As reported in~\cref{tab:vulnerable-fields}, \truncatepercent\% of the analyzed CVEs involved a packet that was truncated to shorter than the expected length.
The caller of~\cref{algorithm:SystematicOrderedGeneration} generates these with modest post-processing: remove bytes from the end of the packet and then update checksums in earlier layers.


\subsubsection{Stateful}

In our CVE study, we found that many CVEs can only be triggered from certain states of a protocol.
We examined existing network testing tools to identify one that can reach many states of a protocol.
We chose the~\packetdrill tool.
It is designed to drive a network stack through the state machine for various protocols~\cite{cardwell_packetdrill_2013}.
It supported two transport-layer protocols (TCP, UDP) and two network-layer protocols (IPv4, IPv6), with a corpus of $>$200 scripts that test different functionalities of the TCP protocol. 

We developed a corpus of 7 Packetdrill scripts that can reach the 7 different TCP states where a packet can be injected (LISTEN, SYN-SENT, ESTABLISHED, FIN-WAIT-1, FIN-WAIT-2, LAST-ACK, CLOSE-WAIT). We had only one UDP script as UDP has no states. We use these scripts as test script templates. As shown in \cref{fig:fig-design-overview}, for each test case and mutation instruction generated, we append the mutation instruction to all template scripts of the protocol being tested. This enables us to test all protocol states with the same input.

\subsection{Implementation}

Our \toolname implementation is 3,426 lines of C/C++ and Python.
We describe pertinent aspects of the implementation.

\subsubsection{Portability}

The purpose of \toolname is to support many \acp{ENS}.
Portability is a priority.
As noted in~\cref{sec:background}, ENS have diverse architecture and semantics. 
We de-coupled the \toolname packet generation from the delivery and evaluation of packets.
PacketDrill generates a combination of socket commands and network packets.
As shown in~\cref{fig:fig-design-overview}, a per-\ac{ENS} Test Agent maps POSIX socket commands to the appropriate behavior on the \acp{ENS}.
This includes both minor naming changes (\eg socket vs FreeRTOS\_socket) as well as more substantial semantic changes (\eg rendering the asynchronous socket semantics of LwIP into synchronous POSIX semantics).
This test agent (server) on the \ac{ENS} receives socket interactions and packets and delivers them to the \ac{ENS}.
The rest of the system is agnostic to the \ac{ENS} under test.

\subsubsection{\pd}

We implemented the network testing tool as an extension of~\packetdrill.
We extended the~\packetdrill grammar to support mutation instructions.
We implemented a packet mutator component in C that, given a packet and a set of mutation instructions, mutates the packet following the instructions.
For example, the instructions might be to change the value of a field, insert an option, and truncate the packet.
We modified~\packetdrill so that it loads the Packet mutator as a shared library and uses it for packet mutation. 
\pd also interacts with our portable bridge component instead of directly with the network.

\subsubsection{Packet Injection}

\toolname uses a virtual network interface (TAP~\cite{noauthor_universal_nodate}) to send mutated packets to Embedded Network Stack. A virtual network interface allows us to inject packets at the lowest layer of the network stack, which simulates the exact same behavior as when the ENS receives the packet from the internet, removing the possibility of false positives. Furthermore, a virtual network interface does not introduce the same network latency that would be introduced by a normal network interface connected to the internet.

\subsubsection{Parallelizing test execution}

Once packets are generated systematically, executing them is an embarrassingly parallel problem.
We decoupled test case generation from execution using the producer-consumer pattern, saturating our servers.

\subsubsection{Deduplicating vulnerabilities}

\toolname systematically generates packets, which may result in many redundant defects.
Our crash monitor analyzes the observed failures and deduplicates them based on the stack trace (line of crash). 

\subsubsection{Linux versions of \acp{ENS}}


Although \acp{ENS} support many boards, they also support Linux as a development environment~\cite{Srinivasan2023LinuxRehosting}.
\toolname uses the Linux versions of the \acp{ENS}.
This enables \toolname and the \ac{ENS} to run on the same machine, enhancing communication between them.
This does introduce the risk that our results mask defects in HW/SW integration on real boards, \eg due to layering issues~\cite{Shen2023NCMAs}.

\section{RQ4: Systematic Testing with \toolname}
\label{sec:evaluation}

We evaluate our systematic testing framework by running \toolname on 4 embedded network stacks. Our evaluation aims to understand the extent our systematic testing approach can uncover packet validation vulnerabilities. Specifically, we answer the following questions.

\begin{itemize}
    \item{\textbf{RQ4.1:}} Can \toolname replicate known vulnerabilities? 
    \item{\textbf{RQ4.2:}} Can \toolname discover new vulnerabilities?
    \item{\textbf{RQ4.3:}} What are \toolnames performance characteristics?
    \item{\textbf{RQ4.4:}} How does \toolname compare to fuzzing?
\end{itemize}

\vspace{0.1cm}
\noindent
\subsubsection{Experimental Setup}
We evaluated N = 1,2,3 and we used a stride that yielded 4-6 values for each field (\cref{algorithm:SystematicOrderedGeneration}).

We used the following servers for our experiments: two 32-core machines (Ubuntu 22.04, Intel Xeon W-2295 CPU@3GHz);
  and
  one 64-core machine (Ubuntu 22.04, AMD EPYC 7543P CPU@2.8GHz).

\subsection{Embedded Network Stack Selection}
We selected 4 \acp{ENS} for our evaluation --- FreeRTOS+TCP, Contiki-ng, PicoTCP, and LWIP.
These stacks from~\cref{sec:vulstudy} had the highest proportion of Network and Transport layer vulnerabilities, suiting them for \toolname.

\subsection{\code{\ensbench}: Vulnerability Dataset Contruction}
\label{subsec:vuln-dataset}


To enable us to answer RQ4.1, we replicated \recreatedvulncount known vulnerabilities in recent versions of 3 selected \acp{ENS} --- FreeRTOS, Contiki-ng, and PicoTCP.\footnote{LwIP had no reported IP/TCP vulnerabilities.} These were selected out of the 14 reported vulnerabilities that affected the IPv4, IPv6, TCP, and UDP protocols in the selected \acp{ENS} layer protocols. 
We skipped 2 vulnerabilities because Packetdrill lacked support for the features they required (IPv6 fragmentation).
We studied their fixing commits to replicate the vulnerabilities and reverted the fix.
Porting the vulnerabilities to the latest version allowed us to have all vulnerabilities in a single build for testing.
\cref{tab:recreated-vulns} describes the CVEs we recreated. 


\begin{table}[h]
\caption{
    CVEs \toolname recreates. The last column indicates dependent fields and kind of changes that expose CVE. Notation: F---set header \textbf{F}ield; O---insert+set \textbf{O}ption; T---Truncate header; Rd---\textbf{R}ea\textbf{d}; Wr---\textbf{Wr}ite.
}
\label{tab:recreated-vulns}
\begin{center}
\begin{small}
\begin{tabular}{p{1.375cm}llc}
\toprule
\textbf{ENS} & \textbf{CVE-ID} & \textbf{Type} & \textbf{Operators} \\
\toprule
FreeRTOS & 2018-16523 & Div-by-zero & 1 (O) \\
         & 2018-16524  & OOB Read & 1 (F) \\
         & 2018-16526  & OOB Write & 1 (O) \\
         & 2018-16601  & Integer underflow & 1 (F) \\
         & 2018-16603  & OOB Read & 1 (T) \\
\midrule
Contiki-ng & 2021-21281 &  OOB Read & 1 (F) \\
           & 2022-36053 & OOB Write & 2 (F, T) \\
\midrule
PicoTCP & 2020-17441 & OOB Read & 2 (F, F) \\
        & 2020-17442  & Integer Overflow & 2 (F, O) \\
        & 2020-17444  & Integer Overflow & 2 (F, O) \\
        & 2020-17445  & OOB Read & 2 (F, O) \\
        & 2020-24337  & Infinite Loop & 1 (O) \\
\bottomrule
\end{tabular}
\end{small}
\end{center}
\end{table}

\subsection{RQ4.1: Replicating Known Vulnerabilities}
\label{subsec:known-vuln}

To evaluate \toolname's ability to expose defects, we ran \toolname on vulnerable versions of FreeRTOS, Contiki-ng and PicoTCP. Our test found all vulnerabilities in the tested stacks as listed in \cref{tab:recreated-vulns} using a maximum of 2 mutations. \cref{tab:recreated-vulns} also shows the mutation types performed on the packet that exposed each vulnerability. These vulnerabilities were triggered by a total of 9 distinct fields. IPv6 extension header length caused 3 while TCP data offset caused 2.

\subsection{RQ4.2: Discovering New Vulnerabilities}
\label{subsec:new-vuls}

To evaluate \toolname's ability to discover new defects, we ran \toolname on recent versions of the ENS listed in \cref{subsec:vuln-dataset}. For FreeRTOS and Contiki, we only ran the tests for IPv4 and IPv6 respectively as that was the only IP version they supported.
\cref{tab:selected-stacks} shows the count of vulnerabilities we found in each of the selected stacks. \cref{tab:new-vulns} describes the various vulnerabilities that we found. In our artifact, we also included the specific scripts that exposed each vulnerability and a detailed description and impact of each vulnerability. 

Contiki-ng and PicoTCP confirmed the vulnerabilities we reported, assigned CVE identifiers, and repaired the vulnerabilities.
We have been unable to establish communication with the LwIP team.

\begin{table}[h]
\caption{
    New vulnerabilities \toolname found.
    Notation: Same as~\cref{tab:recreated-vulns}.
}
\label{tab:new-vulns}
\begin{center}
\begin{small}
\begin{tabular}{cccc}
\toprule
\textbf{ENS} & \textbf{CVE ID} & \textbf{Description} & \textbf{Config.} \\
\midrule
FreeRTOS & --- & \emph{No vuln found} \\
\midrule
Contiki-ng & 2023-34100 & OOB Rd (TCP MSS) & 1 (O) \\
           & 2023-37459  & OOB Rd (TCP flags) & 1 (T)\\
\midrule
PicoTCP & 2023-35847 & Div-by-zero (TCP MSS) & 1 (O)\\
         & 2023-35846   & OOB Rd (TCP fields) & 1 (T) \\
         & 2023-35849   & OOB Rd (IP checksum) & 1 (F) \\ 
         & 2023-35848   & OOB Rd (TCP MSS) & 2 (O, O) \\
\midrule
LwIP     & L1  & OOB Rd (TCP options) & 1 (O)\\
\bottomrule
\end{tabular}
\end{small}
\end{center}
\end{table}

We attempted to evaluate \toolname on commercial \acp{ENS}.
We contacted five vendors of real-time OSes and embedded network stacks: WindRiver (VxWorks), Segger (emPower OS, embOS, emNet), Green Hills Software (GHNet), Lynx (LynxOS), and Sysgo (PikeOS).
All declined to allow us to evaluate on their systems.

\subsection{RQ4.3: Performance Characteristics}
\label{subsec:performance}

\myparagraph{Test Execution Duration}
We measured the execution duration of~\toolname by varying the number of dependent fields (\ie $N$ in~\cref{algorithm:SystematicOrderedGeneration}).
\cref{tab:test-duration} shows the test execution duration on PicoTCP.
For each value of $N$, we executed tests over all supported protocols (IPv4, IPv6, TCP, UDP) using 32 consumer instances. 
Our results in \cref{subsec:known-vuln} and \cref{subsec:new-vuls} show that all reported and new vulnerabilities could be found with only N=1 and N=2 tests.
\begin{table}[h]
\caption{
    Performance results from testing on PicoTCP.
}
\label{tab:test-duration}
\begin{center}
\begin{small}
\begin{tabular}{lrcc}
\toprule
\textbf{Task} & \textbf{\# Test cases} & \textbf{Instances} & \textbf{Time} \\
\toprule
One test case   & 1         & 1     & 0.5 sec \\
N=1 test      & 2,211      & 32    & 2.13 min \\
N=2 test      & 134,296    & 32    & 2.21 hr \\
N=3 test      & 5,303,604   & 32    & 63.17 hr \\
\bottomrule
\end{tabular}
\end{small}
\end{center}
\end{table}

\myparagraph{Coverage Analysis}
We analyzed the coverage achieved by running \toolname on 4 \acp{ENS} compiled with~\texttt{gcov}.

\begin{table}[h]
\caption{
    Table showing the line coverage achieved by different tests when executing \toolname on all the tested stacks.
}
\label{tab:test-coverage}
\begin{center}
\begin{small}
\begin{tabular}{lllll}
\toprule
\textbf{Test} & \textbf{FreeRTOS} & \textbf{Contiki} & \textbf{PicoTCP} & \textbf{lwIP} \\
\toprule
Script 1   & 37.4\% & 25.4\% & 11.3\% & 32.6\% \\
Script 2  & 34.1\% & 24.8\% &  5.9\% & 29.9\% \\
All Scripts  & 51.3\% & 29.6\% & 12.7\% & 40.0\% \\
N=1 tests      & 53.4\% & 33.4\% & 14.8\% & 43.6\%\\
\bottomrule
\end{tabular}
\end{small}
\end{center}
\end{table}

\cref{tab:test-coverage} shows the line coverage achieved executing different tests.
For the Integrated \acp{ENS}, we consider only the coverage of the networking component.
The first two rows show the coverage for two \packetdrill tests with scripts representing different TCP states. The third row represents the coverage achieved when we ran stateful test scripts representing all TCP states. The last row indicates the coverage when we ran a systematic test with N=1. By using test scripts that represent different TCP states, we achieved a significant increase in coverage. The little coverage increase caused by N=1 tests shows that packet validation vulnerabilities exist in codes that are covered by normal executions. 
As shown in \cref{tab:new-vulns}, this N=1 was also sufficient in detecting most of the new vulnerabilities we found.
We could not get very high coverage as the \acp{ENS} contained protocol implementations in other network layers that we don't currently support.

\subsection{RQ4.4: Fuzzing Comparison}

We used the Contiki-ng fuzzing benchmark provided by Poncelet 
\etal~\cite{poncelet_so_2023} to demonstrate that within a time budget, fuzzing is not deterministic in uncovering vulnerabilities. We selected 4 fuzzers from the benchmark (MOpt~\cite{lyu2019mopt}, Intriguer~\cite{cho2019intriguer}, SymCC~\cite{poeplau2020symbolic}, and AFL).
The first 3 had the best results during Poncelet~\etal's evaluations.
AFL is a standard comparison point.
To help the fuzzers, we (1) disabled checksums in Contiki-ng, and (2) augmented the fuzzers' seed set with \toolnames comprehensive set of seed packets. 

\begin{table}[h]
\caption{
    Fuzzing results with the Contiki-ng fuzzing benchmarks after 24 hours.
    Version 1 contains a version of Contiki-ng used by the Poncelet~\etal authors for evaluation.
    It contains the vulnerabilities reported by the authors in their paper.
    Version 2 is the most recent commit on Contiki-ng on GitHub as of May 1st, 2023.
    This version contains 5 vulnerabilities, including 2 detected by \toolname. 
    These vulnerabilities should cause a crash in V2 if triggered.
}
\label{tab:test-contikiBench}
\begin{center}
\begin{small}
\begin{tabular}{lll}
\toprule
\textbf{Metrics} & \textbf{Version 1~\cite{poncelet_so_2022}} & \textbf{Version 2} \\
\toprule
\# paths covered        & 190    & 316 \\
Crashes found (\#)      & 21    & 0 \\
Hangs found (\#)        & 12    & 0  \\
\bottomrule
\end{tabular}
\end{small}
\end{center}
\end{table}

After 24 hours, the second column of \cref{tab:test-contikiBench} shows none of the fuzzers triggered any crash in vulnerable Version 2. 


\myparagraph{Comparison with other network protocol fuzzers}
As noted in~\cref{relatedWork}, there are other network fuzzers, \eg TCPFuzz~\cite{zou_tcp-fuzz_2021} and AFLNet~\cite{pham_aflnet_2020}.
They are not appropriate for the vulnerabilities we studied.
For example, AFLNet targets the application layer of the network stack, while TCPFuzz is concerned with semantic defects on legitimate input.


\section{Discussion}

\myparagraph{\toolname vs Fuzzing vs Static Analysis}
Fuzzing is the most used technique for detecting vulnerabilities. As a dynamic analysis technique, fuzzing provides a low rate of false positives. 
But fuzzing requires significant computing resources to be effective, limiting developers' ability to detect vulnerabilities at development time.

Like fuzzing, \toolname is also a dynamic analysis technique. But unlike fuzzing, it completes and provides guarantees that the known patterns of packet validation vulnerabilities do not exist in the ENS. 

Static Analysis is another widely used approach that succeeds in detecting specific vulnerability patterns. Unlike dynamic analysis, many static analysis techniques consider only specific code sections rather than the entire software and give off a lot of false positives \cite{ami_false_2023}. We briefly ran CodeQL \cite{noauthor_codeql_nodate} on the \acp{ENS} repositories and found that it struggled with inter-procedural cases, failing to find any known or new vulnerabilities we detected.

\myparagraph{Learning from Intra and Inter-product Vulnerabilities}
Our results in \cref{subsec:new-vuls} show that the known vulnerability patterns still exist in \acp{ENS}. We found cases in Contiki-ng where they added regression tests for individual errors but failed to generalize these tests to classes of errors. We recommend that software engineers learn from the individual errors that occur in their software, and prepare generalized test cases that can detect similar errors.

We also found that the same vulnerabilities recur in different software implementations. For example, CVE-2023-35847 (\cref{subsec:new-vuls}) in PicoTCP is the same as CVE-2018-16523 in FreeRTOS. CVE-2023-34100 in Contiki-ng is the same as CVE-2018-16524 in FreeRTOS. 
Anandayuvaraj~\etal already performed preliminary studies on this phenomenon of recurring failures in software engineering \cite{anandayuvaraj_reflecting_2023} and initiated conversations towards a failure-aware software development lifecycle \cite{anandayuvaraj_incorporating_2023, anandayuvaraj2022FailureAwareSDLC}. 
Our findings in this paper further emphasize this need for software engineers to learn from the reported vulnerabilities and failures of other software products.
Furthermore, tools like \toolname can help by ensuring that new vulnerability patterns, discovered in one software product, can be easily detected and fixed in every other similar software product they may exist in.

\myparagraph{Integrating Security Protections to Embedded Firmware}
As shown in \cref{rq1}, memory corruption is the most prevalent class of vulnerability reported in \acp{ENS}. 
In addition to detecting and fixing these vulnerabilities, protection mechanisms could also be implemented to harden the embedded devices and mitigate the impact of exploitations. 
Protection techniques such as stack canaries, Address Space Layout Randomization (ASLR), and No-Execute (Nx) regions exist for regular operating systems which makes vulnerability exploitation difficult. 
Unfortunately, Yu~\etal \cite{yu_building_2022} showed that these security protection techniques are missing in embedded systems.
Prior research \cite{gopalakrishna_if_2022, bodei_measuring_2019, toussaint_machine_2020} identified cost as a factor that limits the integration of security in embedded systems.
Our work further illustrates the importance of integrating these security defenses into embedded systems. Hence, we advocate for further research in developing and deploying cost-effective protection mechanisms in embedded systems.

\myparagraph{Improving Testing Practices for Open-source Software}
As seen in \cref{subsec:test-suite-characteristics}, different \acp{ENS} employ diverse methods and implementations for testing. While many of the test suites had unit tests, the size and robustness of the tests varied across different \acp{ENS}. This suggests the need for a standard test framework for testing similar software systems such as \acp{ENS}.

\myparagraph{Future Works}
We identify the following opportunities for research to improve this work.
\begin{itemize}
    \item \emph{Checkpoint-based Optimization:} Stateful testing involves  driving the ENS to a specific state before injecting the test packet. This introduces significant overhead. The use of a deferred forkserver~\cite{noauthor_more_nodate} does not work on multi-threaded or networked applications. In the future, we hope to explore the use of process checkpointing and recovery~\cite{noauthor_criu_nodate} to optimize the execution of stateful tests.
    
    \item \emph{Smart Test Case Generation:} Our current \toolname design depends on generating packets where all combinations of fields, up to a value $k$, can be mutated at a time. We plan to explore using program and dataflow analysis to  understand which packet fields interact or depend on each other during packet processing and prioritize testing the combinations of these interacting fields. 
    This approach will build on existing research on concolic and hybrid testing that integrates static analysis, dynamic analysis, and symbolic execution to aid vulnerability detection~\cite{stephens_driller_2016, yun_qsym_2018}.
    This improvement will lead to optimizations in the time to run multiple field combinations shown in~\cref{subsec:performance}.
    
    \item \emph{Application to Other Protocols:} Our results show that vulnerabilities in all network protocol implementations share similar patterns. Hence, we have two questions. Would we find similar patterns in the implementations of other protocols? Would vulnerabilities in different implementations of the same protocols or protocol groups (\eg cryptographic protocols) share the same patterns?
\end{itemize}

\section{Limitations and Threats to Validity}

\myparagraph{Limitations of \toolname}
While \toolname is effective in discovering packet validation vulnerabilities, it has the following limitations
\begin{itemize}
    \item Our implementation of \toolname is limited to the protocols supported by Packetdrill (TCP, UDP, IPv4, IPv6, and ICMP). We believe \toolname will work for protocols in other layers as they contain vulnerabilities with similar patterns. 
    \item Due to the different architectures of \acp{ENS}, \toolname requires a distinct Test Agent for each \acp{ENS}. We designed a portability layer that makes it easy to implement a Test Agent for any ENS.
\end{itemize}

\myparagraph{Construct Validity}
We studied CVEs using well-known classifications, minimizing construct-related risks.
To mitigate further, we used inter-rater agreement as a check.

\myparagraph{Internal Validity}
We assessed the testing practices of \acp{ENS} by looking at their test suites. The maintainers of these \acp{ENS} may have other testing processes which we don't know about.

\myparagraph{External Validity}
We mitigate one generalizability concern by
examining multiple ENS of both kinds (integrated and
standalone).
We acknowledge that the CVEs in our study (\cref{sec:vulstudy}) may have been found by a small number of persons using specific techniques. There may be other vulnerability patterns in \acp{ENS} not detected or reported. Nevertheless, the patterns we observed in these CVEs helped us find new vulnerabilities.


\section{Conclusion}
Embedded Network Stacks play an important role in enabling interconnectivity in cyber-physical systems. Vulnerabilities in these stacks can have severe consequences. We conducted the first study of packet validation vulnerabilities in ENS. We studied 61 vulnerabilities in 6 ENSs. Our results revealed the root causes of packet validation vulnerabilities and the packet sequences needed to trigger them. 
We found that detecting many of these vulnerabilities required only simple mutations to the test packet.
We designed \toolname following these findings and evaluated our implementation on 4 ENSs. We discovered \recreatedvulncount known and \newvulncount new vulnerabilities. 
Our results show that appropriate systematic testing techniques can aid the timely and guaranteed detection of specific vulnerability classes. Other non-deterministic dynamic analysis techniques, such as fuzzing, should be deferred until applications have been adequately tested. 
Furthermore, \toolname will help maintainers of ENSs detect these packet validation vulnerabilities before deploying to the public. 

\section{Data Availability} \label{sec:DataAvailability}

Our artifact is available at \url{https://doi.org/10.5281/zenodo.8247917}. 
In it, we provide:
\begin{enumerate}
    \item A spreadsheet containing our analysis of known \acp{ENS} vulnerabilities.
    \item The source code of \toolname and subcomponents.
    \item \ensbench: Dataset of \packetdrill scripts to trigger the known and new vulnerabilities reported in this paper.
\end{enumerate}

\ifANONYMOUS
\else
\section{Acknowledgements}
We are grateful to the reviewers as well as J. Jones, W. Davis, and S. Bagchi for feedback on the manuscript.
We thank P. Doshi and K. Robinson for assistance in data collection. 
J. Davis acknowledges support from NSF \#2135156.
A. Machiry acknowledges support from NSF \#2247686 and DARPA N6600120C4031.
Davis and Machiry acknowledge support from Rolls Royce.

The U.S. Government is authorized to reproduce and distribute reprints for Governmental purposes notwithstanding any copyright notation thereon. Any opinions, findings, conclusions, or recommendations expressed in this material are those of the author(s) and do not necessarily reflect the views of the NSF or the United States Government.
\fi



\raggedbottom
\newpage

\bibliographystyle{IEEEtran}
\bibliography{bib/final-bib.bib,bib/paschal-extra-bib.bib}

\end{document}